\documentclass[iop,apj]{emulateapj}
\usepackage{amsmath,amssymb,amstext,hyperref}

\usepackage[all]{hypcap} %Links go to figures; breaks on deluxetables (use \capstartfalse \capstarttrue to fix it)

\usepackage{aas_macros}
\usepackage{natbib}

\usepackage{enumitem}

\shorttitle{Energy dissipation of energetic electrons in the IGM}
\shortauthors{Alexander A. Kaurov}

\begin{document}

\title{Energy dissipation of energetic electrons in the inhomogeneous intergalactic medium during the epoch of reionization}
\author{Alexander A. Kaurov}
\affil{The University of Chicago}
\email{kaurov@uchicago.edu}

\begin{abstract}

We explore a time-dependent energy dissipation of the energetic electrons in the inhomogeneous intergalactic medium (IGM) during the epoch of cosmic reionization. In addition to the atomic processes we take into account the Inverse Compton (IC) scattering of the electrons on the cosmic microwave background (CMB) photons, which is the dominant channel of energy loss for the electrons with energies above a few MeV. We show that: (1) the effect on the IGM has both local (atomic processes) and non-local (IC radiation) components; (2) the energy distribution between Hydrogen and Helium ionizations depend on the initial energy of an electron; (3) the local baryon overdensity significantly affects the fractions of energy distributed in each channel; and (4) the relativistic effect of atomic cross section becomes important during the epoch of cosmic reionization. We release our code as open source for further modification by the community.

\end{abstract}

\keywords{dark ages, reionization, first stars --- dark matter --- cosmology: theory}
\maketitle

\section{Introduction}

Energy dissipation of high energetic electrons and photons in the IGM is an essential problem when accounting for the effect of Dark Matter (DM) annihilation on the thermal and ionization history of the universe during the epochs of recombination and reionization. Many previous studies on the topic of electron propagation in the IGM were focused on the hard radiation from quasars and a range of energies up to 10 keV. However, various theories of DM, including Weakly Interacting Massive Particles (WIMP) predict that the products of its annihilation consist of much higher energetic particles. Therefore, in this study we explore the energies above 10 keV and up to 10 GeV.

The physical processes that become important for the energy dissipation of the electrons with initial energies between 10-100 keV and 10 GeV are the Inverse Compton (IC) scattering on the CMB photons. Also, due to large time scales, the redshift of photons produced by the IC become a substantial energy drain.

We developed a Monte Carlo code which can account for these additional effects along with the more common ones such as collisional ionization, and excitation, and Coulomb interactions. The objective for this paper is to estimate  the fraction of energy injected into the IGM through different processes. A detailed study of the dissipation of full DM annihilation spectrum after hadronization and its effect on the IGM is out scope of this paper.

The energy range considered in many previous studies \citep{Shull1979,Shull1985h,Dalgarno1999,Valdes2010,Furlanetto2010a} is below 10 keV. In these studies the distribution of energies between atomic processes (ionization, excitation and heat) is explored. We check our results for the consistency with these studies and find reasonable agreement. Since the energy range below 10 keV is well studied we focus on higher energies.

In \citet{Valdes2010} the authors consider higher energies, but do not take into account time evolution and redshifting. In contrast to previous works, we avoid the instantaneous approximation, because the timescales associated with the energies we are considering can be comparable with cosmological. The absorption of energetic photons on cosmological scales is well studied in \cite{Zdziarski1989}.

Time dependent propagation of the energetic electrons and photons during the epoch of recombination is studied in \cite{Slatyer2015} in the context of DM annihilation. Since the universe is very uniform at high redshifts the inhomogeneity of the IGM is neglected in that study. Here we consider lower redshifts (the epoch of reionization), when halos are already formed, and most of the DM annihilation happen within them. Therefore, it is important to consider  the inhomogeneity and local overdensity.

\section{Physical processes}

Our Monte Carlo code starts with an electron (photon) of a given energy. The code evaluates all possible physical processes that the particle can be a part of. These processes can be divided into discrete events and continuous processes. For the discrete events, such as collisional ionization and excitation (and photoionization for photons), the code calculates the probability of the event based on the cross sections and randomly decides whether the particle will interact or not within the current time step. We choose the time step to be small enough, in order to have the probability of the event less than 1\%; therefore, only one or zero discrete interactions happen at each time step. If a secondary particle is formed, i.e. after collisional ionization, the code evaluates its initial energy, and then propagates it independently from other particles with individual choice of a time step.

The continuous processes include electron's deceleration in the plasma and the IC scattering. While these processes are in fact also quantized, we can assume them to be continuous compared to the collisional ionization and excitation due to their higher frequency and lower energy losses per interaction. Therefore, the code integrates the energy losses in each channel during the time step, and then subtracts them from the electron's energy.

A collection of theoretical results on the electron-atom collisional cross sections is available in the convergent close-coupling (CCC) database\footnote{\url{http://atom.curtin.edu.au/CCC-WWW/}}, and there are fits available, for instance, in \cite{Shull1985h}, \cite{Arnaud1985} and \cite{Stone2002}. However, the fits mentioned above are valid only in a non-relativistic case. When energy of an electron exceed $\sim1.5$ MeV, the relativistic effects increase the cross section \citep{Kim2000}. During the epoch of recombination the IC kicks in on energies much lower than $\sim1.5$  MeV (see Figure \ref{fig:ICzones}), and therefore this correction does not play a significant role. However, during the epoch of reionization in moderately overdense regions the relativistic correction becomes noticeable (see \S\ref{subsec:relativity}).

The most important process for energy dissipation of high energetic electrons is the IC scattering on CMB photons. We account for it using Klein-Nishima cross section. Simultaneously with the electron losing its energy we compute the spectrum of the IC photons. The deceleration of an electron due to the interactions with charged particles is calculated with the equations for the Coulomb logarithm given in \cite{Spitzer1962g}.
% Notice, that decelerating electron produces softer spectrum, than the one which has constant velocity\footnote{For instance, the electrons in thermalized plasma.}.

We use photoionization cross sections from \cite{Verner1996}. The elastic and inelastic scattering and pair production cross sections are taken from XCOM photon cross section database \citep{xcom}. Also, we include the cosmological redshifting.

In this study we consider hydrogen and helium atoms, assuming some fixed ionization fraction. Therefore, our model is not fully self consistent, in a sense that we neglect the effect of high energetic particles on the IGM ionization and temperature. We leave a detailed study of the effect on the IGM ionization and temperature throughout the epochs of reionization, and their possible signatures in observations out scope of this letter.

\section{Results}
In this letter we consider only redshift 30 as a representative moment for the epoch of reionization. At this point many DM halos are already collapsed, but the star-producing galaxies are still undergoing formation. Our conclusions hold at all redshifts below ~100. At higher redshifts most of the discussed effects are insignificant due to the homogeneity of the matter distribution.

\subsection{Primary electron}
\label{subsec:primary_electron}

\begin{figure}[!t]
\includegraphics[scale=1.0]{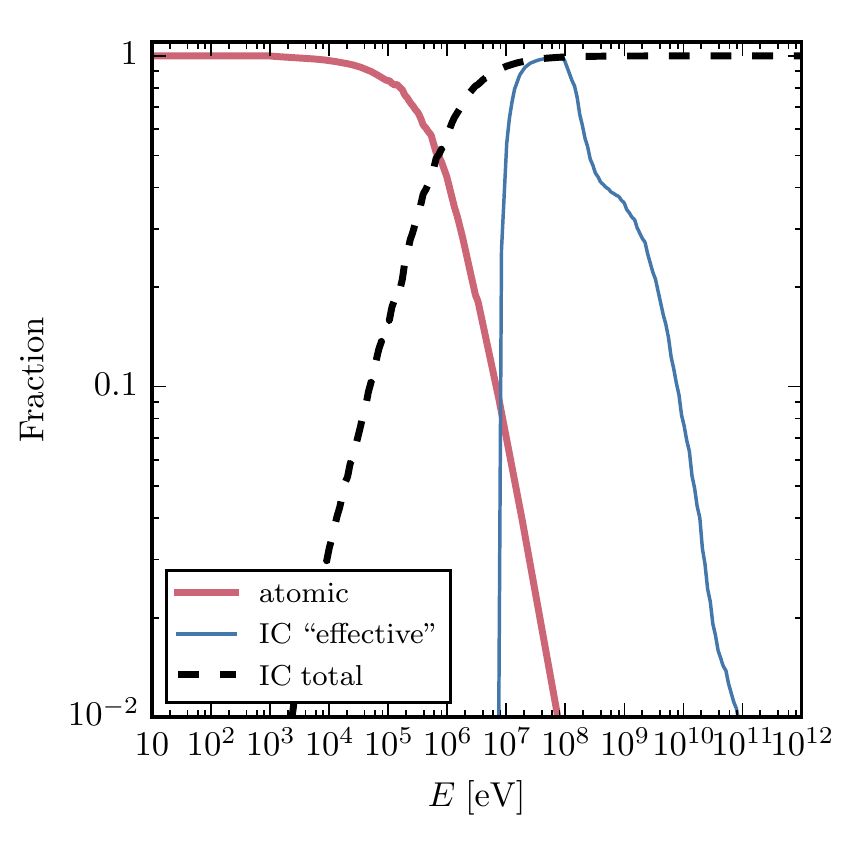}
\caption{\label{fig:channels} The energy fractions dissipated through different channels as a function of the energy of primary electron emitted at redshift 30 in the region with 0 overdensity. \textit{Red thick solid line}: the energy fraction which goes into atomic processes (ionization, excitation and heat). \textit{Dashed black line}: the energy fraction of the IC photons; \textit{blue thin solid line}: the energy fraction of IC photons which interact in any way with IGM before redshift 0.}
\end{figure} 

\begin{figure}[!t]
\includegraphics[scale=1.0]{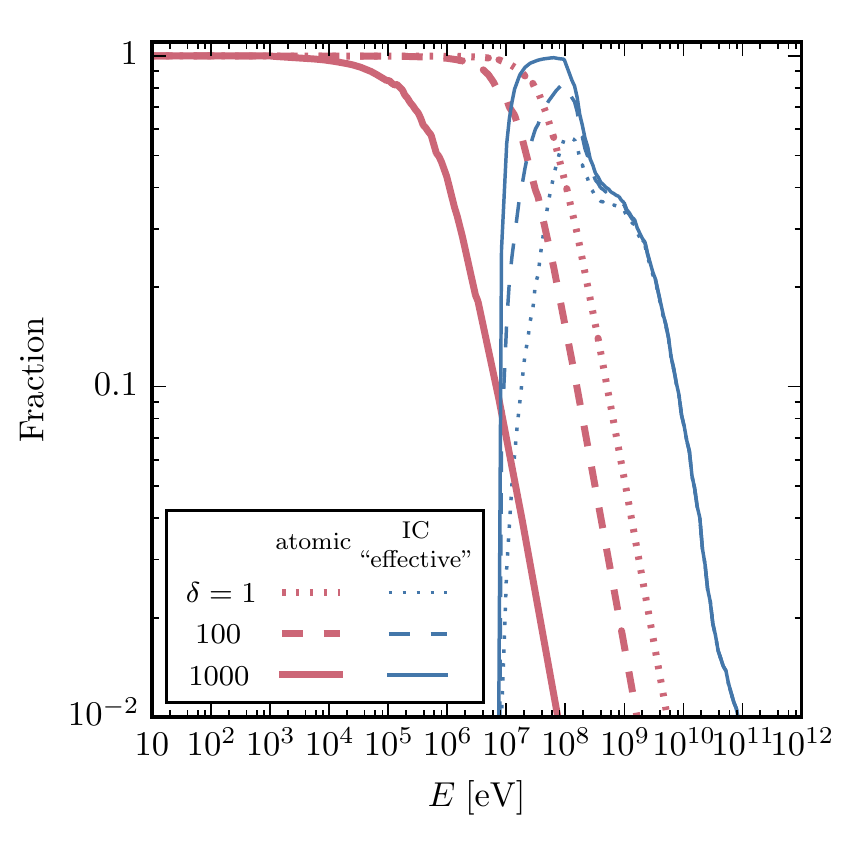}
\caption{\label{fig:vary_delta} The energy fraction which goes into the atomic interactions (\textit{red thick lines}) and  the ``effective'' IC photons (\textit{blue thin lines}) of an electron emitted at redshift 30. \textit{Solid}, \textit{dashed} and \textit{dotted} lines correspond to the local baryon overdensities $0$, $100$ and $1000$. Solid lines are identical to the solid lines in Figure \ref{fig:channels}. Thin blue lines only represent the upper limit (see \S\ref{sec:promary_photon}).}
\end{figure} 

We inject test electrons with various energies at redshift 30 and ambient ionization fractions: HII 1\%, HeII 1\%, HeIII 0\%. The fractions of energy dissipated into ionization and excitation of HI, HeI and HeII, heat, and IC scattering are recorded as functions of time.

For the electrons with energies higher than $\approx1$ MeV, the IC becomes a dominant channel of energy loss. The produced spectrum of the IC photons partially lies under hydrogen ionization threshold, and partially at very high energies where the cross section with atoms becomes small. We discuss the propagation of photons in \S\ref{sec:promary_photon}. Since a photon can easily escape the halo due to a long mean free path and relatively small halo sizes at redshift 30, we assume 0 overdensity and neutral medium while calculating the photon propagation. In reality, things like topology of ionized regions and overdensities might play a significant role. Therefore, with this method we can only estimate the upper limit of effectiveness. 

Having the IC photon spectrum we can subdivide it into three components. First, low energetic photons which do not interact with the IGM. Second, the ``effective'' photons which will interact with the IGM through photoionization, inelastic collisions or pair production. Lastly, the photons which are so energetic that will not interact with the IGM till redshift 0. In Figure \ref{fig:channels} the total fractions of energies deposited into the atomic processes, the IC photons and the ``effective'' IC photons are shown.

Changing the ambient ionization fraction will not affect the IC scattering rate (since it depends only on the energy density of CMB photons). The dependence on the ionization fraction of all other processes is well studied in \citet{Furlanetto2010a} and \citet{Shull1985h}. Therefore, we leave them out scope of this letter. 

The redshift changes the energy of transition to IC regime, but qualitatively the picture remain the same. The rate of IC scattering is proportional to the energy density of photons, therefore it scales as $(1+z)^4$, while atomic processes scale only with the density as $(1+z)^3$. 

\begin{figure}[!t]
\includegraphics[scale=1.0]{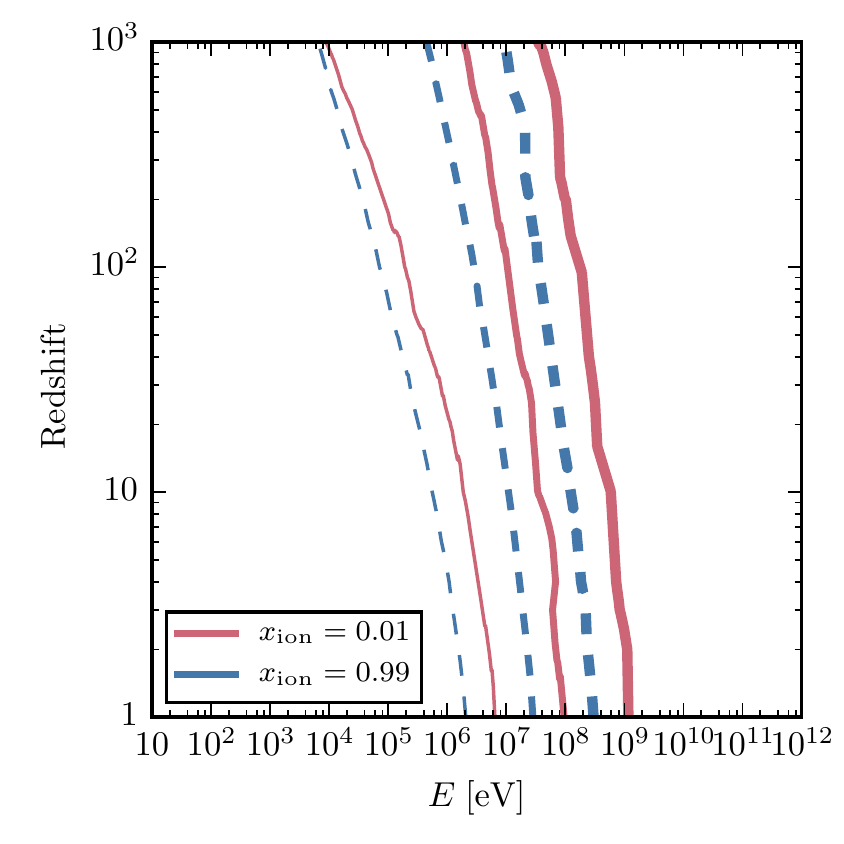}
\caption{\label{fig:ICzones}The energy of a primary electron, at which half of its energy goes into the IC photons and the other half to atomic processes. Solid (red) and dashed (blue) lines correspond to the 1\% and 99\% ionization levels. Thin, medium and thick lines correspond to 0, 100 and 10000 local baryon overdensity.}
\end{figure}

The parameter that might be the most important, especially in the context of the DM annihilation during the epoch of reionization, is the local baryon overdensity. We increase it to $100$ and $1000$, while keeping all other parameters fixed. In Figure \ref{fig:vary_delta} the total fraction of energy that can be potentially absorbed by the IGM is plotted for different local baryon overdensities. The rate of atomic processes increases proportionally to the density, while the density of CMB photons and therefore the rate of IC scattering remain unchanged. It leads to the increase of the transition energy threshold between atomic processes and IC scattering. Consequently, the fraction of the IC photons decreases, followed by the decrease of ``effective'' photons.

In Figure \ref{fig:ICzones} we show the energy of the electrons at which they distribute half of their energy into the IC photons and half into the atomic processes. It shows that in any cosmological environment the IC scattering is significant only for extremely high energetic electrons. For instance, the IC scattering is not important for the electrons produced by X-rays emitted by the hard sources like Quasars.

The spatial scale for the electron energy dissipation is limited by the fact that the ionization cross section, does not go below $10^{-19}\,\mathrm{cm^2}$ for all energies \citep{Kim2000}. Considering $z\approx30$ and ionizing fraction not exceeding 10\%, the mean free path is order of $\sim 500 \mathrm{pc}$. This is close to the galactic scales; however, the galaxies have not been formed at such high redshifts yet. Therefore, we assume that energy dissipation of an electron to be local. In order to properly calculate the propagation of the charged particles in galaxies and, particularly in the Milky Way one has to make assumptions regarding the gas distribution within the disk. The detailed study of the Milky Way is carried out in \cite{Buch2015}.

Also, the time scale at all redshifts (from 0 to 1000s) is smaller than Hubble time \citep{Furlanetto2010a}. Therefore, energy dissipation of a prime electron can be assumed to be instantaneous and the effect on the IGM to be local. However, the IC photons can have large mean free paths and therefore affect the IGM non-locally and with a delay. Figure \ref{fig:vary_delta} shows that the baryon overdensity determines whether the effect on the IGM will be local or not.

\subsection{Primary photon}
\label{sec:promary_photon}

\begin{figure}[!t]
\includegraphics[scale=1.0]{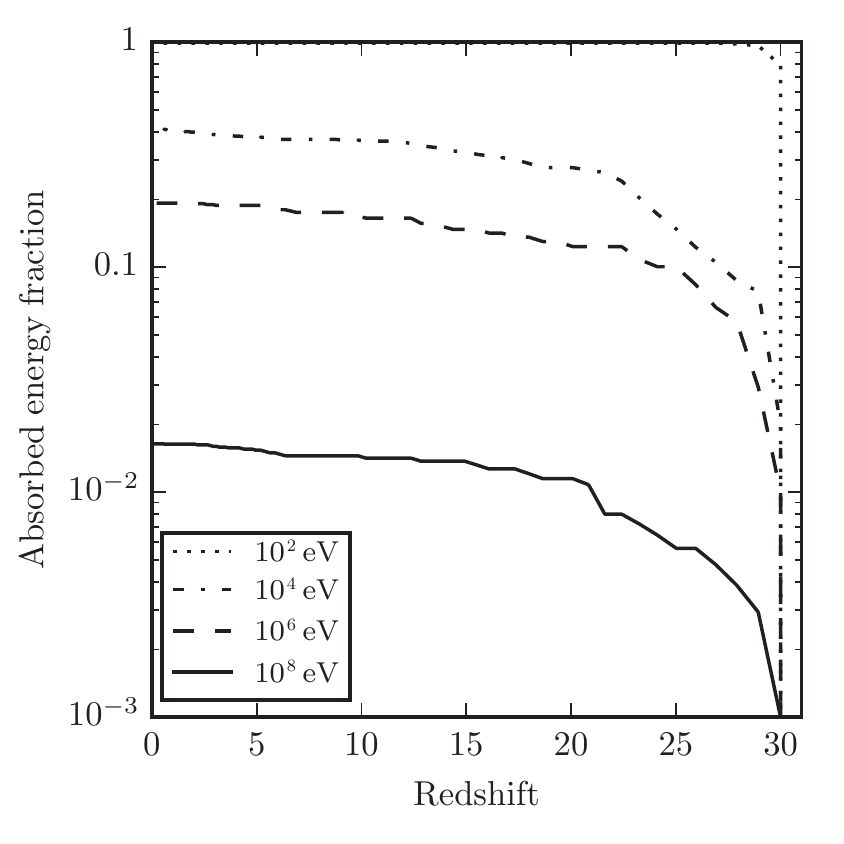}
\caption{\label{fig:photons}The energy fraction of the photons with different initial energies ($10^3$, $10^4$, $10^6$ and $10^8$ eV -- \textit{dotted}, \textit{dot-dashed}, \textit{dashed} and \textit{solid} lines respectively) absorbed or scattered by IGM in any type of interaction. The neutral fraction of IGM is assumed to be 100\% at all redshifts in order to maximize absorption. Therefore these lines represent the upper limit.
}\end{figure}

The IC photons produced by a high energetic electron are also energetic. Therefore we study the energy dissipation of the energetic primary photons. We inject test photons with energies up to $10^8$ eV at redshift 30. The ambient ionized fraction is assumed to be 0 to maximize the absorption rate.

In Figure \ref{fig:photons} the energy fraction of the photons absorbed by the IGM in any process since their emission is plotted. The absorption of photons is not instantaneous even with our assumption of a fully neutral medium. Therefore, the Figure \ref{fig:photons} confirms that the effect of IC photons can be non-local.

\subsection{Ionization of the IGM}
\label{subsec:ionization}

\begin{figure}[!t]
\includegraphics[scale=1.0]{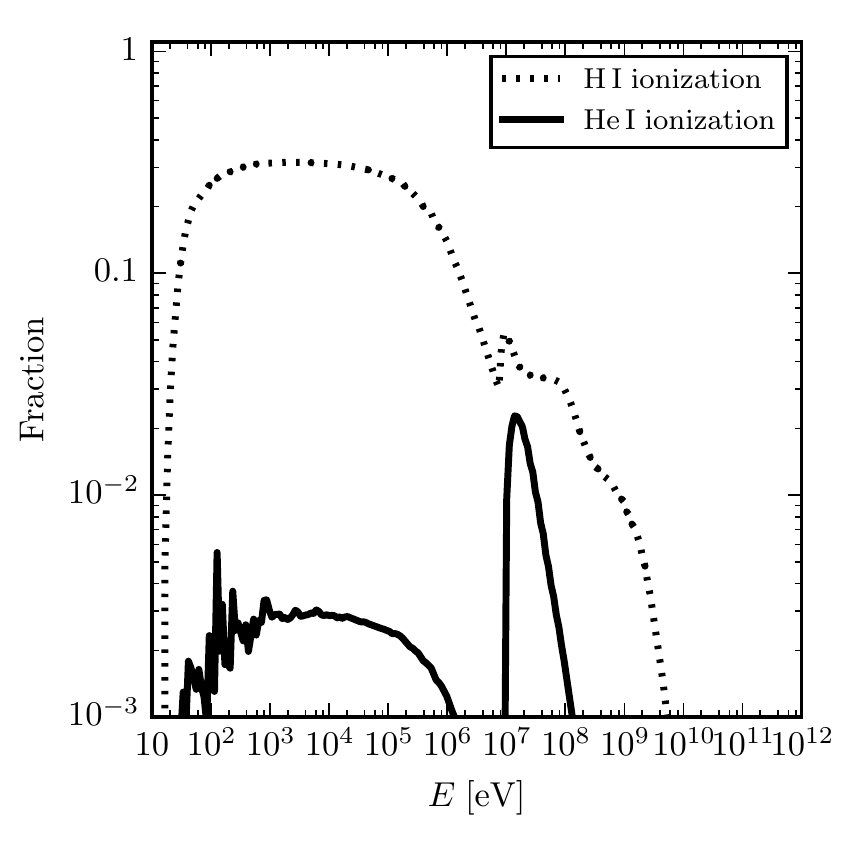}
\caption{\label{fig:ion_channels}The energy fraction of a primary electron which goes into Hydrogen ionization (\textit{dotted line}) and Helium ionization (\textit{solid line}). }
\end{figure}

In the context of the epoch of reionization, the most interesting channel of energy dissipation is the ionization of the IGM. The ionization efficiency of an energetic electrons below 10keV is well studied in \citet{Furlanetto2010a}. Here we added the IC photons, and therefore we consider electrons with higher energies.

We use the same environment parameters as in Figure \ref{fig:channels}. In Figure \ref{fig:ion_channels} the energy fractions which goes to Hydrogen and Helium ionization are plotted. Two main features can be observed.

Firstly, the efficiency of ionization is not uniform across the considered energy range. There is a dip at ~$10^6$ eV associated with the regime where low energetic IC photons are produced, and another dip at ~$10^{10}$ eV, which is associated with a regime when too high energetic photons are produced.

Secondly, the relative fraction of the energy going to Helium is also not constant through out the energy range. At energy around $10^7$ eV the Helium ionization become almost as efficient as Hydrogen ionization. The reason for that is that the IC spectrum produced by the electrons of this energy is peaked near the Helium ionization threshold. At this energy Helium photoionization cross section exceeds Hydrogen cross section, and therefore is ionized more efficiently. 

However, all these calculations are made in assumption of uniform ionization fraction. In reality we expect to have ionization fronts which will complicate the calculation and would require proper radiative transfer models.

\subsection{Relativistic correction}
\label{subsec:relativity}

\begin{figure}[!t]
\includegraphics[scale=1.0]{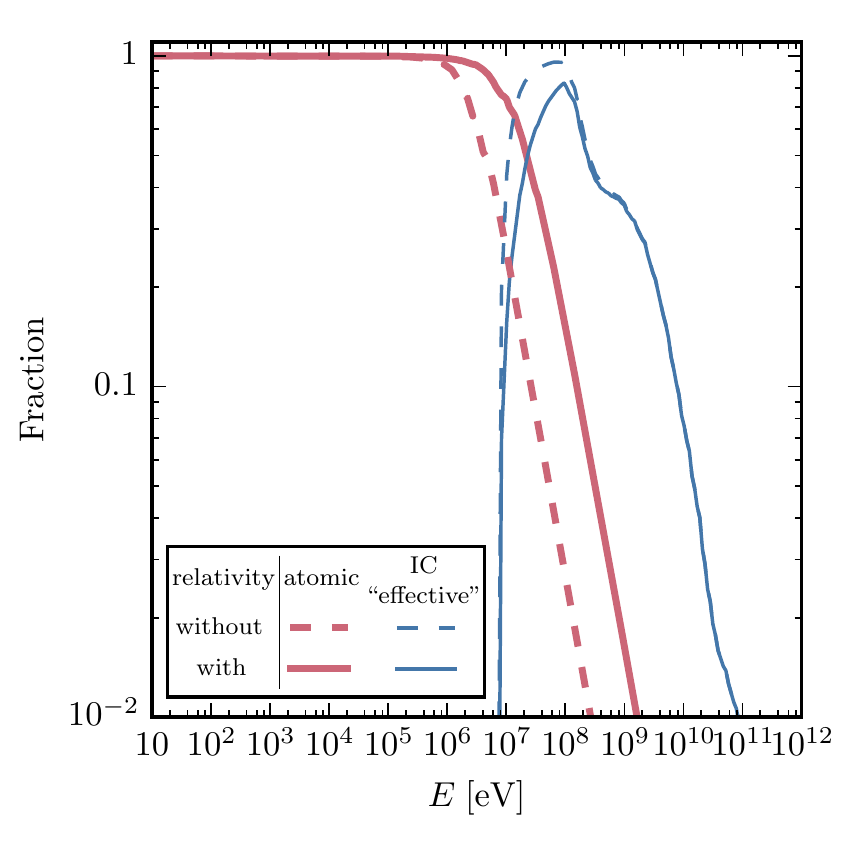}
\caption{\label{fig:relat}The energy fraction which goes into the atomic interactions (\textit{red thick lines}) and  the ``effective'' IC photons (\textit{blue thin lines}) of an electron emitted at redshift 30 and local baryon overdensity $100$. \textit{Solid} and \textit{dashed} correspond to the relativistic and non-relativistic set of cross sections. Solid lines are identical to the dashed lines in Figure \ref{fig:vary_delta}.}
\end{figure}

In Figure \ref{fig:relat} we repeat two curves from Figure \ref{fig:vary_delta} for the overdensity 100, but calculated with and without relativistic correction. The atomic cross sections for the relativistic case are taken from \citet{Kim2000}, and for non-relativistic from \citet{Arnaud1985} and \citet{Stone2002}. In the chosen environment the transition to the IC regime happens around the same energy, at which relativistic effects become important ($\sim1.5$ MeV). At lower redshifts the transition at the same energy would correspond to lower overdensity for the reason mentioned in \S \ref{subsec:primary_electron}. In the energy range $10^6-10^8$ eV the introduced correction can reach a factor of a few. In the other environments where the transition energy is not in the proximity of 1.5 MeV (for, instance during the epoch of recombination) this effect is negligible.

\section{Conclusions}
The study set out with the aim of highlighting the complexity of the energy dissipation of very high energetic electrons during the epoch of cosmic reionization. The results of this study indicate that:
\begin{itemize}[leftmargin=*]

\item There are two components for energy dissipation -- the atomic processes and the IC radiation (\S
\ref{subsec:primary_electron}). The atomic processes affect the IGM locally and almost instantaneously, while the IC photons can travel for a long time before being absorbed (\S\ref{sec:promary_photon}). The channel through which majority of the energy dissipates depends on the initial electron energy. In the presence of DM halos, these two channels would affect the IGM in a dramatically different ways, even if the total energy would be identical. If the electrons have low energies than all the effect is enclosed in the halos; if the energies are high enough, then the impact on the IGM can be global.

\item The ionization rate of different elements depends on the initial energy of the electron (\S \ref{subsec:ionization}). It may lead to the complicated ionization front structure, if those will be driven by the IC photons.

\item The local baryon overdensity affects the energy  distribution between different channels (\S \ref{subsec:primary_electron}). This effect might play a significant role, especially when taking into account that DM and baryon overdensity fields are correlated.

\item The relativistic effect manifests itself in a specific energy range and within it can reach a factor of a few. The environment in which this effect is most pronounced forms only during the epoch of reionization (\S \ref{subsec:relativity}) and not present at the epoch of recombination.
\end{itemize}

This study shows that the energy dissipation of an energetic electron in the IGM interconnects many physical processes. In contrast to the epoch of recombination, the accurate treatment of DM annihilation during the epoch of reionization requires taking into account both the spatial distribution of DM and baryons. However, those are not well known at high redshifts and small scales. This is one of the key issues to consider for future studies.

The code used to perform the presented calculations is released open source as a Python module: \url{http://kaurov.org/codes/radiator/}. It may have important implication in developing more sophisticated models of DM annihilation during the epoch of cosmic reionization.

\acknowledgments
We thank the anonymous referee for critical questions that helped us to significantly improve the original manuscript.

This work was supported by the NSF grant AST-1211190.

\bibliographystyle{aasjournal}
\bibliography{radiator-bib}

\begin{thebibliography}{}
\expandafter\ifx\csname natexlab\endcsname\relax\def\natexlab#1{#1}\fi

\bibitem[{{Arnaud} \& {Rothenflug}(1985)}]{Arnaud1985}
{Arnaud}, M., \& {Rothenflug}, R. 1985, \aaps, 60, 425

\bibitem[{Berger {et~al.}(2010)Berger, Hubbell, Seltzer, Chang, Coursey,
  Sukumar, Zucker, \& Olsen}]{xcom}
Berger, M., Hubbell, J., Seltzer, S., {et~al.} 2010, XCOM: Photon Cross Section
  Database (version 1.5), ,

\bibitem[{{Buch} {et~al.}(2015){Buch}, {Cirelli}, {Giesen}, \&
  {Taoso}}]{Buch2015}
{Buch}, J., {Cirelli}, M., {Giesen}, G., \& {Taoso}, M. 2015, \jcap, 9, 037

\bibitem[{{Dalgarno} {et~al.}(1999){Dalgarno}, {Yan}, \& {Liu}}]{Dalgarno1999}
{Dalgarno}, A., {Yan}, M., \& {Liu}, W. 1999, \apjs, 125, 237

\bibitem[{{Furlanetto} \& {Stoever}(2010)}]{Furlanetto2010a}
{Furlanetto}, S.~R., \& {Stoever}, S.~J. 2010, \mnras, 404, 1869

\bibitem[{{Kim} {et~al.}(2000){Kim}, {Santos}, \& {Parente}}]{Kim2000}
{Kim}, Y.-K., {Santos}, J.~P., \& {Parente}, F. 2000, \pra, 62, 052710

\bibitem[{{Shull}(1979)}]{Shull1979}
{Shull}, J.~M. 1979, \apj, 234, 761

\bibitem[{{Shull} \& {van Steenberg}(1985)}]{Shull1985h}
{Shull}, J.~M., \& {van Steenberg}, M.~E. 1985, \apj, 298, 268

\bibitem[{{Slatyer}(2015)}]{Slatyer2015}
{Slatyer}, T.~R. 2015, ArXiv e-prints, arXiv:1506.03812

\bibitem[{{Spitzer}(1962)}]{Spitzer1962g}
{Spitzer}, L. 1962, {Physics of Fully Ionized Gases}

\bibitem[{{Stone} {et~al.}(2002){Stone}, Y.-K., \& P.}]{Stone2002}
{Stone}, P.~M., Y.-K., K., \& P., D.~J. 2002, J. Res. Natl. Inst. Stand.
  Technol. 107, 327

\bibitem[{{Vald{\'e}s} {et~al.}(2010){Vald{\'e}s}, {Evoli}, \&
  {Ferrara}}]{Valdes2010}
{Vald{\'e}s}, M., {Evoli}, C., \& {Ferrara}, A. 2010, \mnras, 404, 1569

\bibitem[{{Verner} {et~al.}(1996){Verner}, {Ferland}, {Korista}, \&
  {Yakovlev}}]{Verner1996}
{Verner}, D.~A., {Ferland}, G.~J., {Korista}, K.~T., \& {Yakovlev}, D.~G. 1996,
  \apj, 465, 487

\bibitem[{{Zdziarski} \& {Svensson}(1989)}]{Zdziarski1989}
{Zdziarski}, A.~A., \& {Svensson}, R. 1989, \apj, 344, 551

\end{thebibliography}

\end{document}